\documentclass[twocolumn,showpacs,preprintnumbers,aps,amsmath,amssymb]{revtex4}
\usepackage{graphicx}% Include figure files
\usepackage{dcolumn}% Align table columns on decimal point
\usepackage{bm}% bold math
\begin{document}
\title{\it Displacement power spectrum measurement of a macroscopic 
optomechanical system at thermal equilibrium}
\author{A. Di Virgilio$^1$, S. Bigotta, L. Barsotti$^1$, S. Braccini$^1$, C. Bradaschia$^1$, G.
Cella$^1$, V. Dattilo $^2$, M. Del Prete$^1$, I. Ferrante$^3$, F. Fidecaro$^3$,
I. Fiori$^1$, F. Frasconi$^1$, A. Gennai$^1$, A. Giazotto$^1$, P. La Penna$^2$,
 G.Losurdo $^4$, E. Majorana$^5$, M. Mantovani $^6$, F. Paoletti$^{1,2}$,
 R. Passaquieti$^3$, D. Passuello$^1$, F. Piergiovanni$^7$, A. Porzio$^8$, P. Puppo$^5$,
 F. Raffaelli$^1$, P.
Rapagnani$^5$, F. Ricci$^5$, S. Solimeno$^{8,9}$, G. Vajente$^{1,10}$, F. Vetrano$^7$}
\affiliation{$^1$INFN, Sez. di Pisa, Pisa, Italy \\
$^2$ EGO, European Gravitational Observatory, Cascina (Pi)\\
$^3$ Universita' di Pisa, Italy\\
$^4$ INFN Sezione di Firenze, Sesto Fiorentino, Italy\\
$^5$ Universit\`{a} di Roma1, and INFN-Roma1, Roma Italy\\
$^6$ Universita' di Siena, Italy\\
$^7$ Universit\`{a} di Urbino, Urbino, Italy\\
$^8$ Coherentia, CNR-INFM, and CNISM Unit\'a di Napoli\\
$^9$ INFN, Sez. di Napoli, Universit\`{a} di Napoli\\
$^{10}$ Scuola Normale Superiore, Pisa}
\date{\today}
\begin{abstract}
The mirror relative motion of a suspended Fabry-Perot cavity 
is studied in the frequency range
 $3-100$ $Hz$.
 The experimental measurements presented in this paper, have 
 been performed at the Low Frequency Facility,
 a high finesse optical cavity $1$ $cm$ long suspended to a 
  mechanical seismic isolation system identical to that one 
  used in the VIRGO experiment. 
  The measured relative displacement power spectrum is compatible with
   a system at thermal 
equilibrium within its environmental.
In the frequency region above $3$ $Hz$, where seismic noise 
contamination is negligible, the measurement distribution is 
stationary and Gaussian, as expected for a system at thermal equilibrium. 
Through a simple mechanical 
model it is shown that: applying the fluctuation dissipation theorem
 the measured power spectrum is reproduced below $90$ $Hz$ and noise 
 induced by external
 sources are below the measurement. 
\end{abstract}
\maketitle
\section{Introduction}
Thermal fluctuations of mechanical systems are considered the most 
relevant limitation of ground based interferometers for gravitational 
waves detection in the low frequency region, where several gravitational 
wave signals are expected\cite{teo}. The statistical behavior is well
 described by the Fluctuation Dissipation Theorem (FDT), which states a general relationship
 between the response of a given system to an external disturbance and 
 the internal fluctuation of the system in the absence of the
 disturbance. Such a response is characterized by a response function or 
 equivalently by an admittance, or an impedance \cite{kubo,fluctuationdiss,saulson}.
The test masses for gravitational wave interferometer are complex 
mechanical systems where contributions to the thermal noise come from 
different elements of the structure: the suspension wires of the mirror, the mirror bulk 
and the mirror coating, which is considered one of most severe thermal noise source 
for the present and near future detectors, just to mention a few of them. 
Measurements devoted to the evaluation of different dissipation mechanisms 
and of the parameters describing them, are often performed on resonance. 
Just a few experimental apparatus have been conceived to measure 
the thermal noise spectrum out of resonance and in the frequency region 
above $100$ $Hz$ 
\cite{gonzales,yamamoto,heidmann,numata}.
The Low Frequency Facility (LFF) is the test bench on top of which the 
Italian R\&D program of the VIRGO experiment is based on. Its main purpose 
is the relative displacement measurement of two suspended mirrors forming a 
high finesse Fabry-Perot cavity suspended from a seismic isolation system and the 
study of the thermal noise spectrum in the region 
starting from $10$ $Hz$\cite{okinawa}. 
This is made possible by a Superattenuator 
(SA), very similar to the ones installed in VIRGO,
 at the INFN Pisa laboratory. Care has been used in designing and 
assembling this experiment in order to 
keep the experimental apparatus as close as possible to the VIRGO's suspensions.
Moreover, since the seismic 
noise reduction to the payload level is very large, the LFF represents 
also an independent benchmark of the VIRGO suspension behavior in the low 
frequency region. 
Recently, data collected at the LFF has shown the presence of an optical spring due 
to the radiation pressure noise of the system
\cite{prl}. 
Within the frequency band $3-90$ $Hz$ this thermal-noise dominated system is 
unique as well as the presence of the optical spring acting between two 
mirrors of the cavity. 
The first section of this paper shortly describes the 
experimental apparatus. An introduction of the mechanical model reproducing
 the experimental measurements is presented in the second section. 
 Within the third section  is focused on the evaluation of the 
 main noise coming from external sources in LFF. 
 In the last section, before the conclusions, 
 a general overview on the measured spectra 
 is given, paying particular attention to the main characteristics of 
 the thermal noise driven mechanisms. Then the absorption coefficients 
 found fitting the data with the model
 are discussed. 
\section{The Experiment set-up and Data Acquisition}
\begin{figure*}[tbp]
\includegraphics[width=12cm]{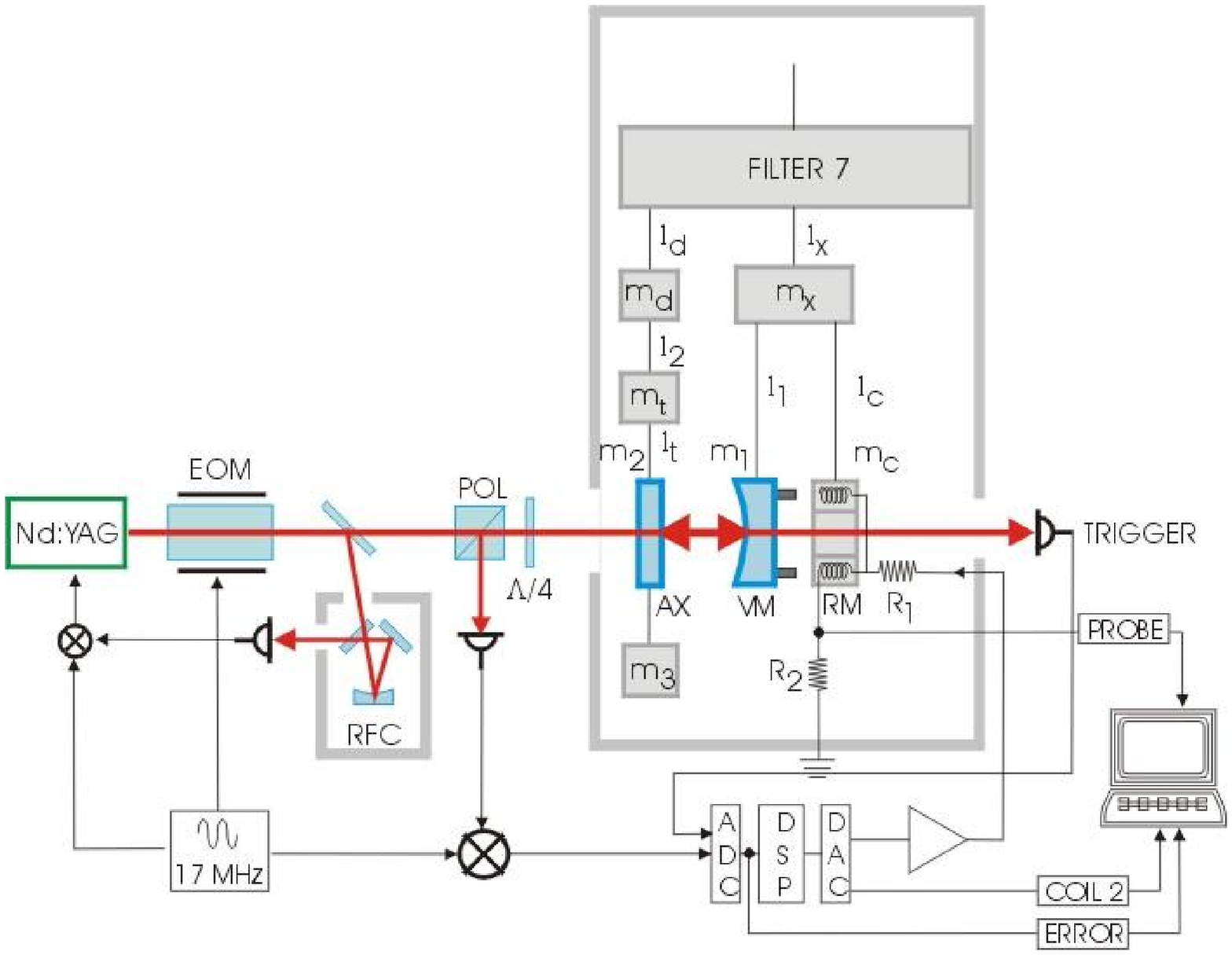}
\caption{Sketch of the experiment set-up from Filter7.
 The optical circuit, and the control loop are shown; gray boxes
 underline the components under vacuum.}
\label{fig:eletmec}
\end{figure*}

Last stage of the experimental apparatus is sketched in fig. 
~\ref{fig:eletmec} 
while a more detailed description can be found in references 
\cite{okinawa}.
 The reflectivity, transmission and loss ($R_i$, $T_i$, $A_i$ with $i=1,2$ respectively) 
 of the two mirrors used in our set-up are given in table \ref{table_optics} 
together with the nominal cavity finesse and the laser input power. RFC (reference cavity) 
and the suspended mirrors (enclosed in figure in grey boxes) are in vacuum. 
\begin{table}
\begin{center}
\begin{tabular}{|c|c|c|c|}
\hline
$R_1 $ & 0.9991 \\ \hline
$R_2 $ & 0.9999 \\ \hline
$A_1 and A_2 $ & 0.00001 \\ \hline
$T_1 $ & 0.0009 \\ \hline
$Nominal Finesse$ & 6300\\ \hline
$Pin$ & 200 mW \\ \hline
\end{tabular}
\end{center}
\caption{Optical parameters of the cavity}
\label{table_optics}
\end{table}
The Finesse, estimated from the transmission profile of the un\-controlled cavity,
 is about $5500$ $\pm 1500$. The static radiation 
pressure force is about $5$ $\mu N$.
The suspension system adopted to insulate from seismic noise the high 
finesse $1$ $cm$ long Fabry-Perot cavity is a single SA chain, equal to the ones
 of 
VIRGO interferometer test masses[4]. The suspension and the cavity are in vacuum inside a tank. 
The curved mirror is 
a $25$ mm diameter mirror embedded in a large steel cylinder to form its
 holder and it plays the role of the VIRGO test mass (VM). The flat mirror 
 of the cavity (AX, auxiliary mirror) is hung, by means of an 
 independent three-stage suspension, to the last mechanical seismic filter 
 of the chain called Filter7. It includes an intermediate mass $m_d =71.72$ kg 
and an additional smaller clamp of the AX suspension wire,  $m_t=0.08$ kg. 
The control of the longitudinal motion is 
 done by acting only on the VM mirror using the $2$ coil-magnet pairs. 
 This actuation technique is identical to the one implemented in the 
 VIRGO interferometer \cite{SA,GIOV,LS} where the coils are screwed on the reference 
 mass (RM) and the magnets are glued on the back side of the mirror holder. 
Fig. ~\ref{fig:eletmec} shows the cavity, the input beam, 
the longitudinal control loop scheme 
and the acquired signals. The laser beam, phase modulated at $17$ 
$MHz$ and independently 
frequency stabilized on a 15 cm rigid ULE reference cavity, 
is injected into the cavity. The reflected power, deviated by the polarizer, 
is detected by a photodiode and demodulated by the mixer (Pound-Drever-Hall scheme, P-D-H),
 with a scheme analogous to the one used for the 
VIRGO pre-stabilization circuit, which should provide a laser 
frequency stabilization at the $mHz/\sqrt{Hz}$ level
 \cite{VIRGObook}. The linear zone of the read-out signal is about $10^{-10}$ $m$.
The feedback control loop is based on a Digital Sygnal Processor (DSP) developed for the
 VIRGO suspension system control. The cavity signal is amplified and then sent
  to a $16$ bit ADC while the DSP  takes care of the signal filtering 
  to control the stability of the longitudinal electromechanical loop. 
  The filtered signal is sent to four DAC ($20$ bit) channels and then to the 
  four coil drivers. The error signal is acquired with a LabView program  
  using a $16$ bit ADC.\\
  Mixer output, coil voltage, filtered by an anti-aliasing filter at $3.4$ $kHz$, are the relevant
  signal acquired during measurement, with acquisition rate of $400$ $Hz$.
   They are indicated in fig. ~\ref{fig:eletmec}
  as ERROR, COIL2 and PROBE. 
\section{Measurements and mechanical model}
\begin{figure*}[tbp]
\includegraphics[width=12cm]{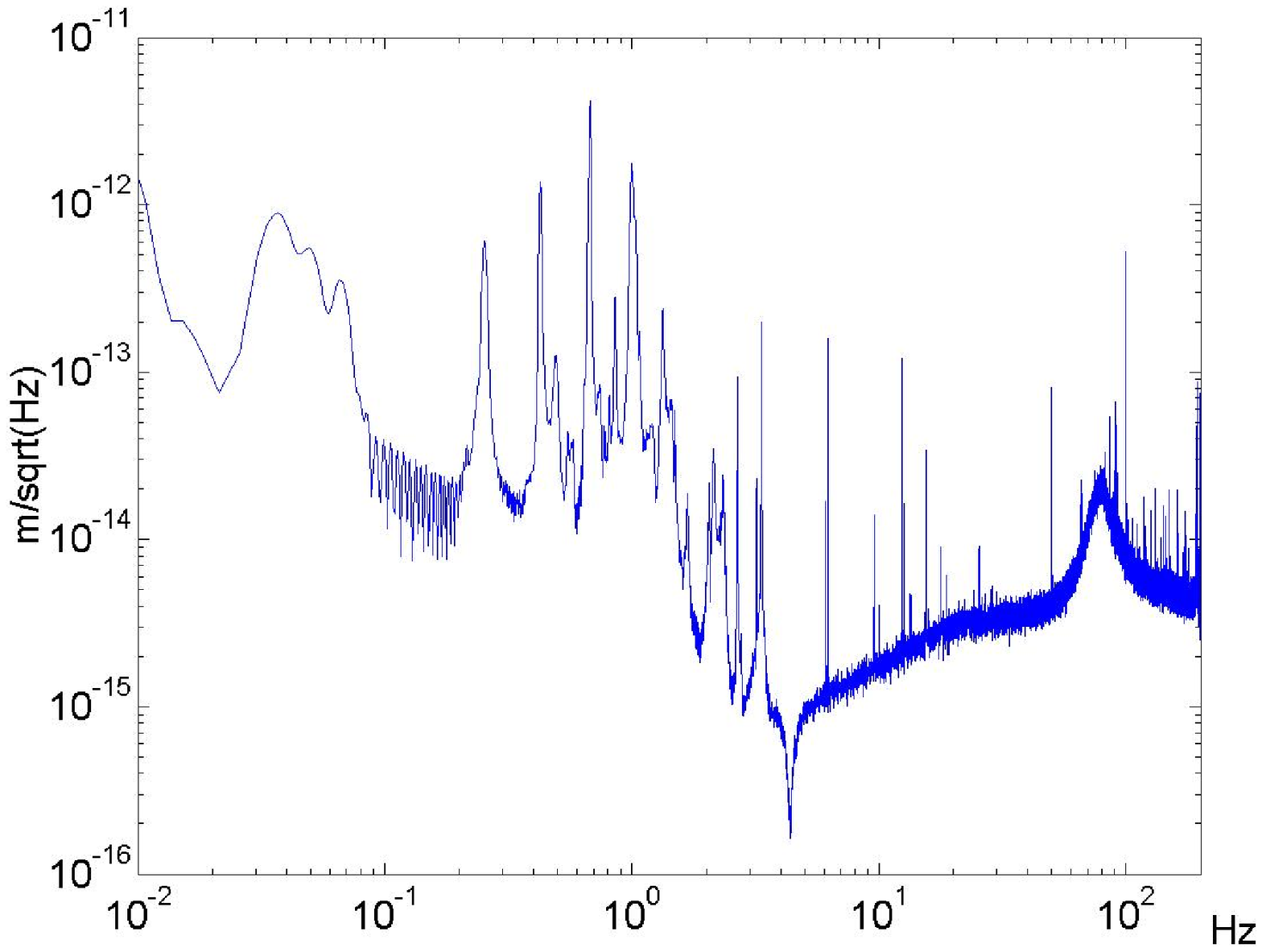}
\caption{Typical power spectrum of the relative motion of
the suspended cavity, while the loop is active.}
\label{fig:misura0}
\end{figure*}

\begin{figure*}[tbp]
\includegraphics[width=12cm]{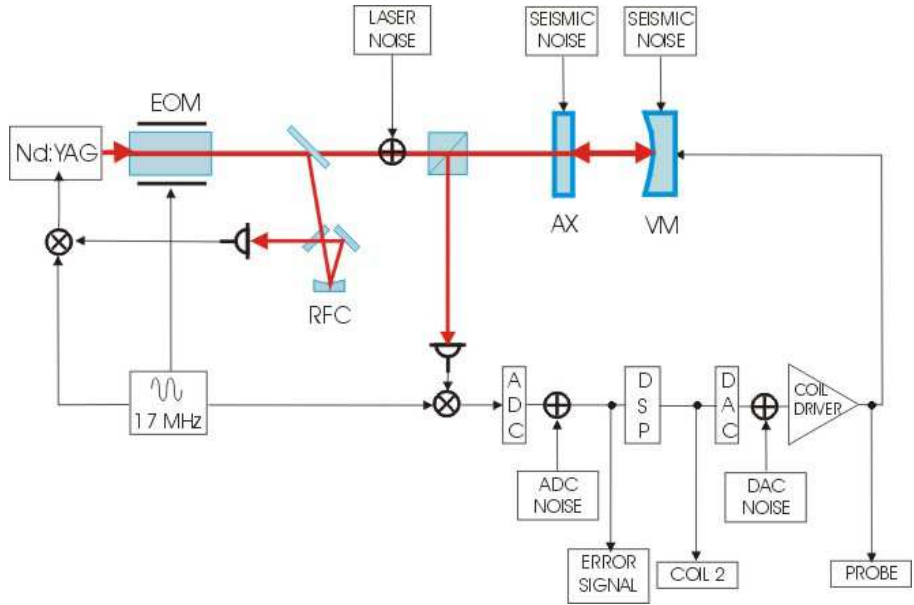}
\caption{Scheme of the control loop, which includes the main noise sources}
\label{fig:feed}
\end{figure*}

In fig.~\ref{fig:misura0} the power spectrum of the two mirrors relative motion suspended 
to the SA chain 
is shown. This plot has been obtained analysing a collected data stretch $1$ hour long. 
  A typical structure of the SA chain excited by the seismic noise is visible below $3$ $Hz$ 
while at higher frequency a few peaks with high mechanical quality factor (Q) are present. 
In particular, the broad peak coming from the optical spring at about $90$ $Hz$ is well visible.
To better understand the behavior of the LFF experimental set-up,
 a one-dimensional model with mass-less wires and point-like masses has been 
developed. Although the SA chain adopted to isolate the optical components from seismic 
noise is a complex system, the mechanical model is confined to the study of the dynamical 
system formed by the two mechanical branches hung to the Filter7 (see fig. ~\ref{fig:eletmec}). 
Each branch includes one mirror of the optical cavity: the first one (AX) is made of four 
masses suspended in series while the second one (VM) is composed by a simple pendulum from 
which two elements are attached in parallel.
In Table \ref{table_suspensions} the mechanical parameters of the LFF model are collected.
The feedback loop circuit has been included within the model taking into account the filtering
 calculation performed by the DSP as well as the coil driver impedances (see
 fig.~\ref{fig:eletmec}). 
 The magnetic actuator coupling constant is $\alpha= 3$ $mN/A$
  while the typical gain of the optical
  read-out is $1.56\times10^{10}$ $V/m$. The transfer function of the
   function running in the DSP 
  consists of the sum of two pieces: one has a real pole at $0.5$ $Hz$ 
  (with gain $13$ at frequency zero), and
 the other a real zero at $30$ $Hz$ (with gain $5$ at $750$ $Hz$).
 The low frequency feedback gain is of the order of $10^5-10^6$, 
  large enough to reduce the large motion of the mirror from a 
  few microns down to $10^{-11}$ $m$, well
  inside the linear region of the P-D-H signal.  
\begin{table}
\begin{center}
\begin{tabular}{|c|c|c|c|}
\hline
\textbf{Masses} & kg & \textbf{Wires} & mm \\ \hline
$m_{d}$ & 71.72 & $l_{d}$ & 1100 \\ \hline
$m_{t}$ & 0.08 & $l_{t}$ & 250 \\ \hline
$m_{2}$ & 0.296 & $l_{2}$ & 500 \\ \hline
$m_{3}$ & 2.5 & $l_{3}$ & 300 \\ \hline
$m_{x}$ & 80 & $l_{x}$ & 1130 \\ \hline
$m_{1}$ & 27.61 & $l_{1}$ & 700 \\ \hline
$m_{c}$ & 64.14 & $l_{c}$ & 500 \\ \hline
\end{tabular}
\end{center}
\caption{Mechanical characteristics of the LFF systems.}
\label{table_suspensions}
\end{table}
The system is affected by different noise sources 
entering through different 
dissipation mechanism, fig. ~\ref{fig:feed} shows the control loop, the 
acquired signals and the external noises entering in the system.
This implies that the model should account for the possibility
of changing the entering position of the noise sources
(physically resembling the different sources affecting different stage of the 
electro mechanical system). In particular, the critical points
are suspension, mirror, electronic noise of the DSP and coil driver.\\
The thermal noise of a mechanical system can be calculated using the
fluctuation dissipation theorem \cite{fluctuationdiss,kubo}. 
The theorem defines a relation between the 
frequency distribution of thermal noise in a system and the 
mechanical admittance $Y$, which is 
the ratio between the resulting velocity and the applied force (i.e. the transfer function of 
the velocity as a function of the applied force):
\begin{equation}
Y(\omega)=\frac{v}{F}
\end{equation}
Specifically, the displacement spectral density is given by:
\begin{equation}
x^2(\omega)=\frac{4 k_B T}{\omega^2}\Re(Y(\omega)) 
\end{equation}
where $k_B$ is the Boltzmann's constant, $T$ the temperature and $Y$ the 
admittance, $\omega=2\pi\nu$, where $\nu$ is the frequency.
In our case $x$ is the distance from the resonance point of the cavity, as well as the
 relative motion of the two mirrors of the cavity, also called
error signal of the feed-back longitudinal control loop. The model takes into
account 
the rotational degree of freedom that can be excited by an off-axis beam,
necessary since the beam hits the mirror few millimeters far from the center. 
This has been done
by adding a single equation to the model, thus considering the flat mirror as a single 
oscillator of given momentum of inertia and a proper frequency around $3$ $Hz$, as 
experimentally observed.
This approximated models held because the proper frequencies of multi-pendulum suspension  
are well below the range analyzed
in this context.
 The elastic constant of the rotation is $0.13$ $N/rad$, with a
inertial moment of the mirror $3.2 \cdot 10^{-4}$ $kg m^2$, the misalignment
 between the vertical rotational axis of the AX mirror and 
the point where 
the radiation pressure force is applied is $4$ $mm$.\\
To take into account the losses we add inside the equation 
of motion a term proportional 
to the velocity (viscous damping) or 
we multiply the
spring constants by $(1+i\phi)$ (structural damping), where $\phi$ is
the loss angle (usually considered constant and smaller than $\phi=10^{-4}$)\cite{norna,geppo}.\\
In the model only two dissipation terms have been considered:  
 pure viscous
damping acting on the longitudinal degree of
freedom, and the other acting on the rotational one of the AX mirror.  This  simple 
choice is based on the fact that the AX pendulum has larger thermal noise, since
its mass is $100$ times lighter than the other. Moreover  
we can assess which dissipation mechanisms is dominant by analyzing 
the different frequency behavior, under the assumption that
the absorption parameters are constant with the frequency.\\
Assuming that the control loop does not introduce noise \cite{ritter},
the model evaluates the thermal noise contribution to the relative motion $x$ of 
the mirror of the cavity in the following way:
\begin{itemize}
\item in the open loop case, the mechanical impedance is 
evaluated by applying two equal and opposite forces to
the two mirrors of the cavity, in the model the optical spring is described by a standard stiffness
\item the thermal noise power spectrum of $x$ is evaluated by using F-D-T
\item the stochastic forces are evaluated by the thermal noise power spectrum of $x$ and the model
\item the closed loop motion, to be compared with the measurement, is evaluated by
applying the  thermal stochastic forces to the mirrors in the closed loop equations
\item free parameters are the absorption coefficients and the optical spring stiffness, 
which are derived by fitting the data with the model.
\end{itemize}
\begin{figure*}[tbp]
\includegraphics[width=12cm]{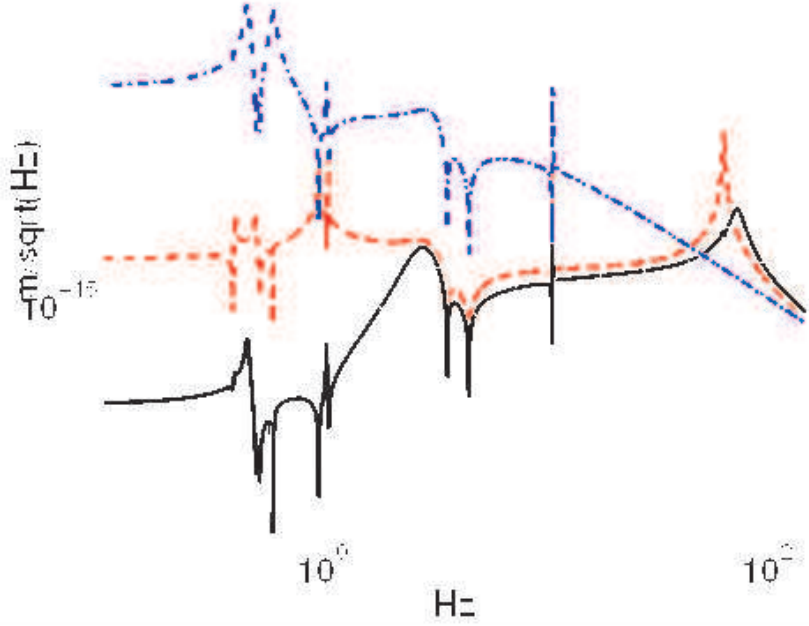}
\caption{Thermal noise power spectrum in the three cases: open feedback loop without optical spring (dot-dashed line)
 open  feedback loop with optical spring (dashed line), 
closed  feedback loop with
optical spring (continuus line), typical LFF parameters are used.}
\label{fig:thermal_all}
\end{figure*}
Figure ~\ref{fig:thermal_all} shows the thermal noise spectra in the following cases:
open loop with and without optical spring and the closed loop with optical spring (using the typical
feed-back parameters used in the runs which have been analyzed).

\section{External noises contributions}
\begin{figure*}[tbp]
\includegraphics[width=12cm]{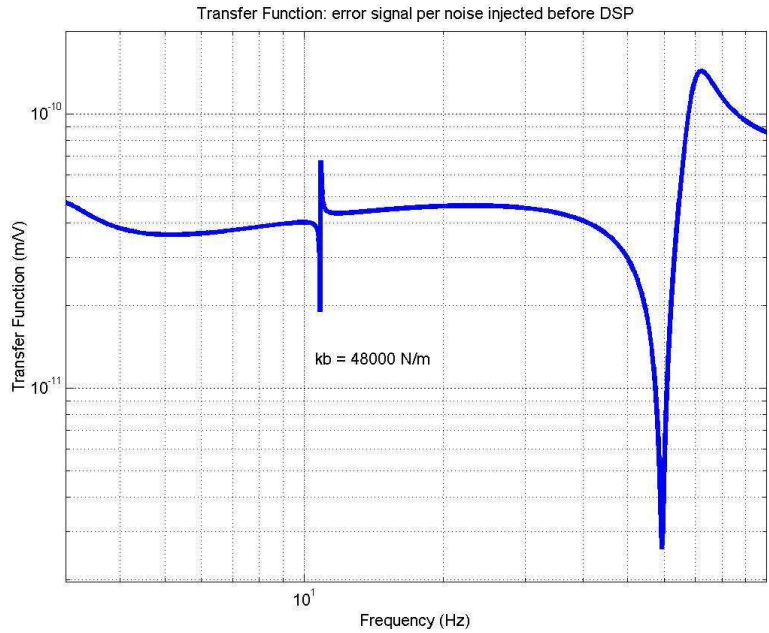}
\caption{ PROBE Output versus frequency computed in the model  by injecting  a white noise before the DSP.  }
\label{fig:tfmn}
\end{figure*}

\begin{figure*}[tbp]
\includegraphics[width=12cm]{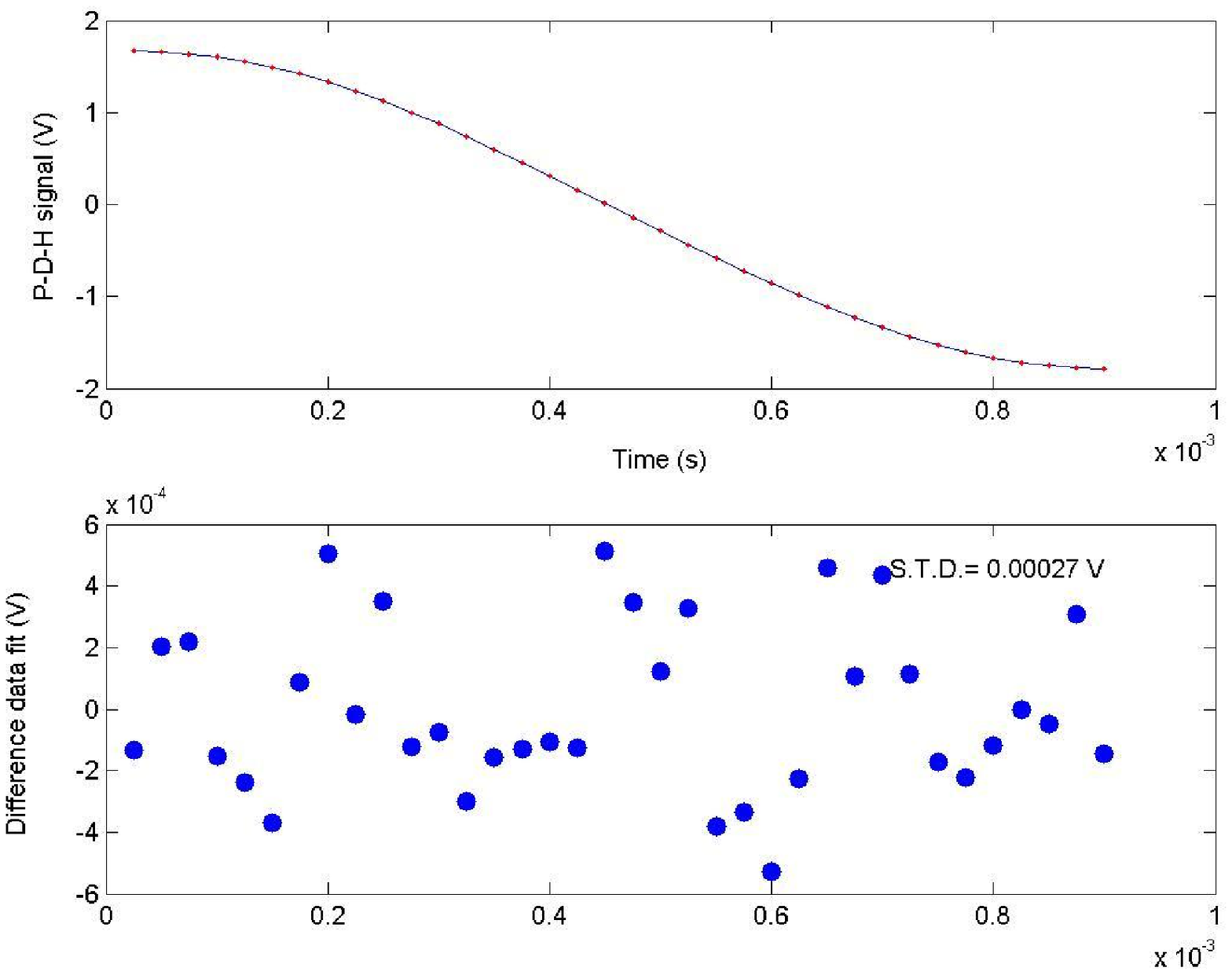}
\caption{up: P-D-H signal around the resonance, feed-back not active, over imposed is the the polynomial fit of the recorded
points up to order 9.\\
down: difference measured points polynomial fit, which is a good measure of the noise coming from the laser, frequency and
amplitude noise }
\label{fig:freq}
\end{figure*}

 Few peaks are evident in the spectrum: the $12$ $Hz$ peak is related to the existence of the small clamp $m_t$,
  all the
 others have not been identified. Some of the peaks are not stationary; comparing the data of
 two different time periods of a single 
acquisition run, it has been checked that the peak heights are enhanced while
 the noise floor remain unaffected \cite{okinawa}.
 For this reason those peaks are not crucial for the comprehension of the noise floor; 
 they are related even to the excitation of degrees of freedom orthogonal to the longitudinal one ,
 or  through the input beam which is 
in air, or to spurious electromagnetic couplings.\\
 In the following the measured power spectrum will be compared with all the external noise sources.
Thanks to the fact that the measurement is stationary, all not stationary noise sources (as
Doppler shift due to the jitter of the injection beam, 
excited by seismic noise or by fluctuations of the air pressure induced by sound) are ruled out.
As earlier mentioned, the different noise sources enter the system at different points, see fig. ~\ref{fig:feed}.
 While seismic noise
enters only through the suspension, electronic noise comes in through several ports: 
the laser (frequency and amplitude jitters),
the mixer, the photodiode, the ADC, the DAC and the coil drivers.
Direct measurement of
the electronic noise has shown that the main contributions come from ADC and DAC.
The electronic noise of the DAC and the driver cannot produce the measured spectrum.
 In order to produce a power spectrum similar to the measured one, the noise of the DAC should be
  $2\times 10^{-4}$
 $V/\sqrt{Hz}$, which is more than a factor
  $100$ higher that the measured DAC noise.\\
  Signals are acquired just
 before the DAC (COIL2) and after at the level of the coil (PROBE, current flowing in the coil);
 the coherence of the signal COIL2 with the signal PROBE is very high, over $95\%$ above $10$ $Hz$:
   this confirms  that the DAC noise is negligible.\\ 
Figure ~\ref{fig:tfmn} shows the ratio between a noise 
injected before the DSP filter (as noise from the laser, mixer, photodiode and ADC noise)
 and the error signal; fig. ~\ref{fig:tfmn}
shows that a white noise would not reproduce the data in the frequency band $3-100$ $Hz$. 
In order to match 
the $80$ $Hz$ peak a noise with a spectral density of $1.5\times10^{-3}$ $V/\sqrt{Hz}$ should be injected.
on the other hand the ADC noise spectrum is white and its value is a factor $100$ lower.\\
The frequency jitter of the Nd-YAG laser is one of the most important noise sources. 
For this reason, as we said before, we set -up  a laser frequency stabilization loop 
which should limit the fluctuations at the $mHz/\sqrt{Hz}$ level. However, since we  
haven't an independent reference cavity to perform a direct measurement of the residual noise, 
we have a strong evidence that this noise source does not contribute significantly to the noise floor of our set up.  
First all we notice that, in order  to reproduce the measured power spectrum level around the 
optical spring resonance, we need  a noise frequency jitter of  $\sim350$ $Hz/\sqrt{Hz}$. 
This value is just a factor $5$  below the measured laser frequency jitter in absence of stabilization. 
Moreover, we were able to evaluate an upper limit for the integrated  
frequency jitter with the following independent measurement.
Before the runs, for calibration purposes, the P-D-H reflected signal was recorded, 
at $40$ $kHz$ acquisition rate, 
with the cavity out of lock. 
Each point taken in the linear region of the P-D-H signal has an error due to
 the frequency and amplitude jitter of the laser, and any other possible noise source, integrated 
from $0$ to $20$ $kHz$. Among all the recorded data an event has been selected, with over
 $30$ points in the linear
region: the
cavity was moving slowly and with constant speed. Thirty six points have been recorded close to  the linear region or the P-D-H
signal, the data have been fitted with a polynomial up to ninth order.  The difference between the points and the polynomial
is a estimation of the noise affecting the measurement integrated from zero up to $20$ $kHz$.
 The standard deviation of this noise is $2\times10^{-4}$ $V$,
incompatible with a power spectrum density of $1.5\times10^{-4}$ $V/\sqrt{Hz}$. In figures
 ~\ref{fig:freq} the data, the residuals of the fit are shown.
 \\
 A detailed analysis of the seismic noise contamination can be found elsewhere \cite{sisma}. Here we recall 
 that the contamination of seismic noise at $10$ $Hz$ is below $10^{-15}$ $m/\sqrt{Hz}$ for the VIRGO interferometer
 test masses.

\section{Fit results and Discussion}
 
\begin{figure*}[tbp]
\includegraphics[width=12cm]{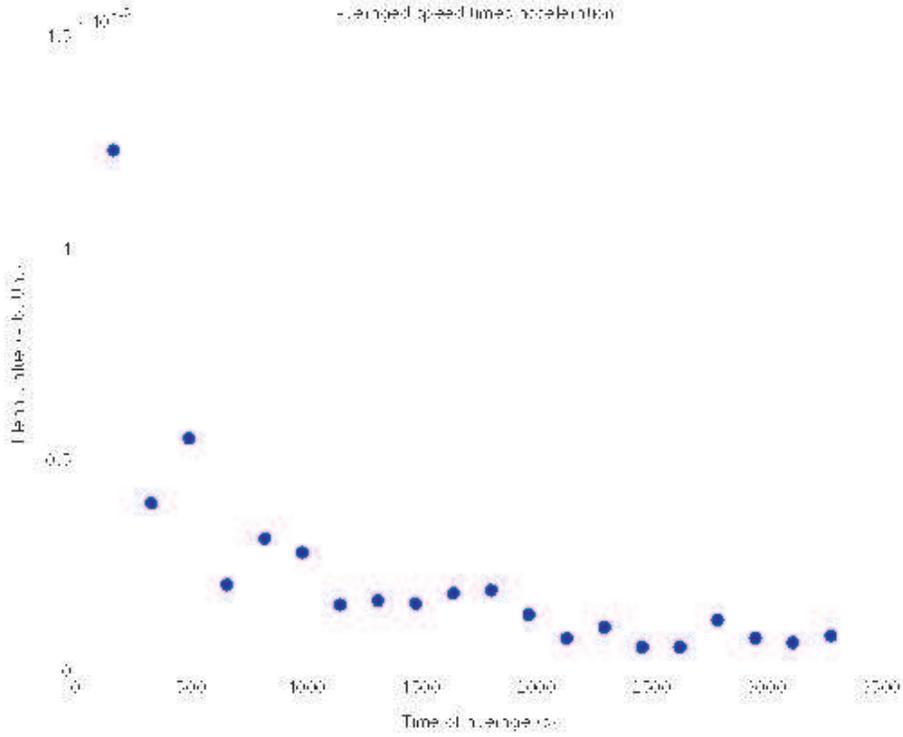}
\caption{Speed times acceleration averaged over time intervals of increasing length}
\label{fig:speedacc}
\end{figure*}

The study presented here is based on a set of data collected in different working conditions 
but with similar control loop gain and the test performed confirms
 compatibility of the data with the thermal noise prediction.
As mentioned in a previous paragraph the most important signal used in our analysis is
 that related to the cavity de-tuning (closed loop error signal) obtained by reconstructing 
 the signal output of the DSP (COIL2 in fig.~\ref{fig:eletmec}). 
 Indeed, this signal is acquired after a  two poles Butterworth filter, 
 which selects the frequency range $10-200$ $Hz$.  The region 
of the spectrum below $3$ $Hz$ is dominated by the seismic noise and it has been cut
 off by an other high pass filter applied to the data. \\
By means of the Horde software routine of the MatLab software package, we checked that  
 data follow the Gaussian statistical distribution, while   the data stationary has been checked by 
 looking at  the time-frequency spectrograms of the data.\\
According to the reference \cite{kubo} an independent check of the data stationary for 
a thermal noise dominated system can be evaluated averaging the product between 
the speed and the acceleration of the observed system ($<v_x\times a_x>$, $v_x$ and $a_x$ 
are the speed and the acceleration).
 This parameter should vanish if the system is thermal noise dominated.
The fig. ~\ref{fig:speedacc} shows the distribution of the mentioned parameter (the average 
of the product between the speed and the acceleration of the system) averaged on 
increasing time interval and for frequencies below $100$ $Hz$. 
The speed and the acceleration have been obtained differentiating the displacement measurements. 
\\
As stated by the Boltzman's law, the distribution of the speed $(v_x)$  for an oscillator 
 excited by the thermal noise should be:
\begin{equation}
f(v_x)=(\frac{m}{2 k_b T})^(\frac{1}{2})\times e^{-\frac{m v_x^2}{2 k_b T}}
\end{equation}
where $m$ is the oscillator mass, $k_b$ is the Boltzmann's  constant and T
the absolute temperature.
Moreover, the variance $\sigma$ of the distribution $f(v_x)$ for a free oscillating system 
(not feedback controlled) is defined by the mechanical parameters. 
Our measurements have been performed with an active feedback loop which does not influence
 the distribution of the data but changes the absolute value of the variance
  itself. In principle the variance $\sigma$, for a closed loop case, could be
   reconstructed integrating the expected power spectrum all over the frequency band. 
   The comparison between the measured spectrum and the predicted one, 
is the most reliable method to prove that the data are thermal noise dominated.
Thus, we checked that the energy associated with the mirror AX,
integrating the power spectrum between $3$ and $100$ $Hz$, is below $k\times T/2$ ($T \simeq 298$ K).

\begin{figure*}[tbp]
\includegraphics[width=12cm]{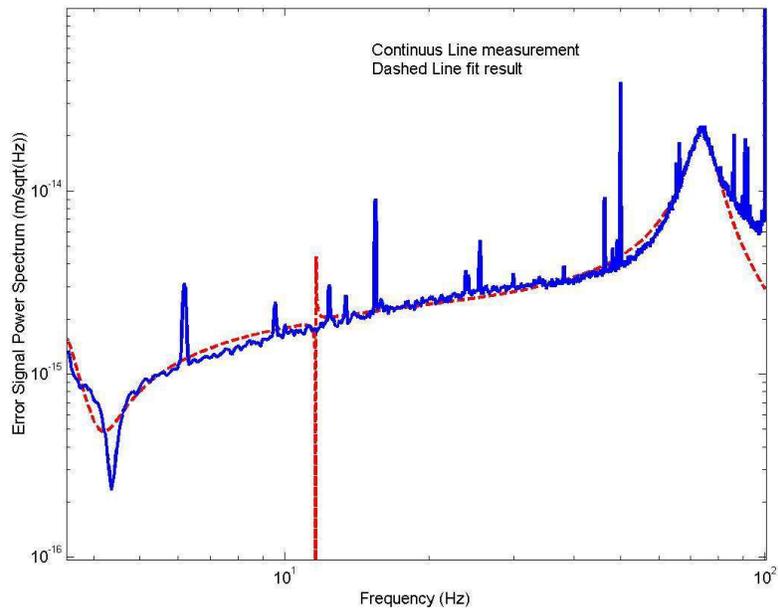}
\caption{Measured power spectrum, $10$ $mHz$ frequency resolution, compared
with the thermal noise estimated by the model, assuming an optical gain
$1.56\times10^{10}$ $V/m$, $k_b= 55000$ $N/m$, and the typical working conditions of the present set of runs, the losses,
two parameters constant in frequency, are associated
to the AM longitudinal and rotational degree of freedom, their fitted values are $5.8$ $\frac{1}{s}$ 
for the longitudinal and $6.5$
$\frac{1}{s}$}
\label{fig:mismatch}
\end{figure*}
\begin{figure*}[tbp]
\includegraphics[width=12cm]{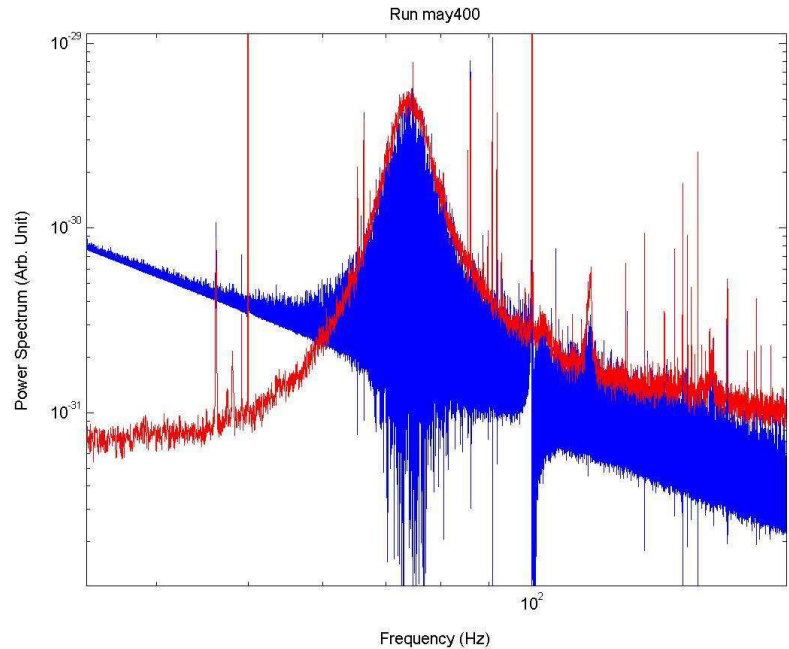}
\caption{Measured power spectrum compared with the power
 spectrum reconstructed by evaluating the admittance from the speed autocorrelation function 
 and applyimng the Fluctuation Dissipation Theorem.}
 \label{fig:kubo}
\end{figure*}

Kubo\cite{kubo} states that for a system at the thermal equilibrium the admittance $Y$ can be 
evaluated by the Fourier transform integrated from zero to infinity of the autocorrelation function of the speed. 
From the knowledge
of $Y$ it is possible to reconstruct the power spectrum from the fluctuation dissipation theorem. 
Let $x_F$ be the 
displacement filtered by the feedback loops and by the filters of the acquisition system.
 The following relation held for the power spectrum
 of $x_F$:
\begin{equation}
S_{x_{F}x_{F}} =\frac{2}{\omega^2} \Re\lbrack\ \int_0^\infty <\frac{d x_F(t+\tau)}{dt}\frac{d x_F(t)}{dt}>e^{i\omega t}\,dt\rbrack
\end{equation}\\
In fig. \ref{fig:kubo} the $S_{x_{F}x_{F}}$ is evaluated both directly from the data 
and applying the above equation. The agreement is good in particular in the region of the
 optical spring resonance and at higher frequencies.
At lower frequency the stretch of data is too short to be used for deriving a significant 
estimation of  the power spectrum on  the base of the  formula reported above.
The good agreement  shown is not sufficient to demonstrate
 that our measurements are thermal noise dominated. It represents just an efficient cross-check 
performed using  uncalibrated and unfiltered data.     
\\
To evaluate the thermal noise contribution to the data we used our model to best fit the data. 
 First all, we derived the absorption coefficient $\gamma_l=5.8$ $1/s$, associated to the
 longitudinal motion of AX, by fitting the data selected around
the optical spring resonance $65-80$ $Hz$, while the other $\gamma_\theta=6.5$ $1/s$ is found by fitting the data in the
 interval $3.5-6$ $Hz$,where there is the anti-resonance.\\
In the figure ~\ref{fig:mismatch} we show the measured spectrum and the  thermal noise contribution estimated by
best fitting the data with the model for an optical gain $1.56\times10^{10}$ $V/m$ and $k_b = 56000$ $N/m$. 
Similar results have been obtained by analyzing the data of independent runs.
\\ It has been also checked that present measurement is not compatible with the 
structural damping.
As it is shown in figure \ref{fig:mismatch}, the result of the fit and
the data well agree below $90$ $Hz$, at higher frequency the higher order modes are relevant, and the model cannot
reproduce the data.\\
 The mirrors are attached through wires, with diameter $\phi=300$ $\mu m$, which in principle should be
treated in the model as continuous system. The family of transversal modes of the wires (violin modes) starts around $100$ $Hz$. 
It has been checked with
the model that higher internal modes of the mirrors and violin modes of the wires,
change the slope of the power spectrum above $100$ $Hz$, and increases the level at lower frequency.
The present set of data shows the power spectrum below $200$ $Hz$  and it is impossible to further constrain the 
 model by adding higher order modes, since the high frequency resonance cannot be identified.
As far as the frequency region below $3$ $Hz$ is concerned, we said before that  
we did not include in the model the super attenuator oscillation modes. However,  a more detailed model 
would not improve the analysis, since this part of the spectrum
 is contaminated by seismic noise. 
\\
We notice that the absorptions coefficients are about $10$ times larger than those measured in VIRGO.
  The LFF apparatus is not as performing as VIRGO from the point of view of the losses due by mechanical dissipation. 
In fact, the LFF apparatus has not been constructed with all the care used 
for the VIRGO antenna; for example
 three wire loops are attached to the AX mirror, the wires diameter is
 rather large ($300$ $\mu
m$), the two wire loops holding the mirror are attached to a motor used to align horizontally the cavity. Due
to the fact that the difference in length of the two wires loop, only $3$ $mm$ apart, was causing a large vertical
misalignment, it has been necessary to change the length of one of the two wires 
loops by using a clamp, which is attached directly to the wires of one of the
  two loops. This clamp consists of two pieces and a screw,  
  which changes the distance between the two pieces, and accordingly the effective 
length of one of the two wires
loop, thus making it possible to bend the AX
mirror.\\
In the model the losses are associated to the AX mirror motion, 
and the higher order mechanical modes of the system are not taken into account.
Indeed we checked that  higher order modes can contribute in decreasing the
absorption coefficients.\\ 
Moreover, we notice that the the values of the two $\gamma$ coefficients resulting from the data fit 
depend also on the absolute calibration: in particular they decrease when  the optical gain is increased.
For example, assuming a optical gain $6\times10^10$ $V/m$, 
the fit gives  $\gamma_l=0.13$ $1/s$ and $\gamma_\theta=1.06$ $1/s$.
The calibration of the LFF apparatus so far relies on
measurements taken
when the cavity feedback is off (open loop condition).  
 The actuator system is calibrated  applying a slowly increasing voltage to the coil
 drivers and we count the number of free spectral ranges
 transmitted by the cavity. In practice, the optical read out is calibrated by looking at the
 Pound-Drever signal; the two sidebands $34$ $MHz$ apart,
  set the scale for the absolute calibration.
  In order to improve the calibration accuracy of the measurement, it would be necessary 
  to have an independent system pushing the mirror of a know amount during  the data taking based for example 
on the radiation pressure effect of an independent laser light, amplitude modulated at a
 fixed frequency impinging on the AX mirror.
   would move the AX mirror of $10^{-14}$ $m$, at $20$ $Hz$, if the optical spring is $64000$ $N/m$.
 With the use of this calibrator it would be possible  
to directly excites the system for a direct measurement of the absorption mechanisms, on resonance and outside resonance.
\section{Conclusions}
A $1$ $cm$ Fabry-Perot cavity is suspended using a super attenuator chain equal to the ones used for the VIRGO antenna.
This cavity has been locked for several hours, and the data analyzed
off-line.The output  signal exhibits a statistical behavior a displacement power spectrum compatible
 with the condition of a system  at its thermal equilibrium.
 We developed  a simple opto-mechanical model which include the various noise sources of the system   
 and we evaluated the thermal noise contribution applying  the fluctuation dissipation
 theorem.
Two viscous dissipation coefficients $\gamma_l$ and $\gamma_\theta$, constant in frequency, associated with the 
longitudinal and rotational motion of the lighter test mass of $0.350$ $kg$, are used in the model
 to fit the data;
the fit results gives $\gamma_l=5.8$ $1/s$ and $\gamma_\theta=6.5$ $1/s$. Different runs give similar results. 
In particular the measured power spectrum
is reproduced in the region $3-90$ $Hz$ under the hypothesis that it is thermal noise dominated.
We checked that the measured power spectrum cannot be produced by external noise sources, as noise of 
the laser and electronics
and that the 
other  noise source in the control loop  are not contributing significantly.
\section*{Aknowledgements}
We would like to thank all the technicians of the Pisa-INFN sections, who have contributed to the
construction of the LFF experiment:
 R. Cosci, C. Magazzu', A. Di Sacco, A. Del Colletto,
F. Mariani,and M. Iacoponi. We thank M. Percimballi and M. Iannone of Rome, P.
Dominici of Urbino and S.Di Franco of Florence.
Special thanks goes to M. Ciardelli, R. Macchia, F. Nenci and A. Pasqualetti, now working for EGO.
%--------------------------------------------------------------------

\end{document}